\documentclass[opinion]{sigirforum}

\usepackage{booktabs}
\newcommand{\fff}{{Ferrante et al.}}
\newcommand{\method}[1]{\mbox{{\small\sf{#1}}}}
\newcommand{\metric}[1]{\method{#1}}
\newcommand{\RR}{\metric{RR}}
\newcommand{\AP}{\metric{AP}}
\newcommand{\NDCG}{\metric{NDCG}}
\newcommand{\RBP}[1]{\metric{RBP}_{#1}}
\newcommand{\Prec}{\metric{Prec}}

\newcommand{\var}[1]{\mbox{\emph{#1}}}

\usepackage{xcolor}

\newcommand{\ansa}{``({\emph{a}})''}
\newcommand{\ansb}{``({\emph{b}})''}
\newcommand{\ansc}{``({\emph{c}})''}
\newcommand{\question}[3]{\medskip\par\noindent{{\bf{Question #1}}~ #2}{({\emph{a}}) #3; or ({\emph{b}}) it is not appropriate to compute an average;
or ({\emph{c}}) insufficient information has been provided in the
question to allow the choice between options {\ansa} and {\ansb} to
be made.}}

\newcommand{\UNI}{{\cal{U}}}
\newcommand{\nU}{u}
\newcommand{\Uprof}{{\UNI}_{\mbox{\scriptsize{prof}}}}
\newcommand{\Urr}{{\UNI}_{\mbox{\scriptsize{RR3}}}}
\newcommand{\Uprec}{{\UNI}_{\mbox{\scriptsize{Prec3}}}}
\newcommand{\UFFF}{\var{equi-interval}}
 
\begin{document}
\title{Categorical, Ratio, and Professorial Data:\\
The Case for Reciprocal Rank}

\authors{
\author[ammoffat@unimelb.edu.au]{Alistair Moffat\\(ORCiD: 0000-0002-6638-0232)}{The University of Melbourne}
	{Australia}
}

\maketitle 
\begin{abstract}
Search engine results pages are usually abstracted as binary
relevance vectors and hence are categorical data, meaning that only a
limited set of operations is permitted, most notably tabulation of
occurrence frequencies, with determination of medians and averages
not possible.
To compare retrieval systems it is thus usual to make use of a
categorical-to-numeric effectiveness mapping.
A previous paper has argued that any desired categorical-to-numeric
mapping may be used, provided only that there is an argued connection
between each category of SERP and the score that is assigned to that
category by the mapping.
Further, once that plausible connection has been established, then
the mapped values can be treated as real-valued observations on a
ratio scale, allowing the computation of averages.
This article is written in support of that point of view, and to
respond to ongoing claims that SERP scores may only be averaged if
very restrictive conditions are imposed on the effectiveness mapping.
 \end{abstract}

\section{Introduction and Background}
\label{sec-intro}

Researchers often work with datasets of categorical or ordinal
observations, on which only a restricted range of operations may be
undertaken.
In particular, search engine result pages (SERPs), which are
typically abstracted as $k$-element binary vectors that tabulate the
relevance or not of the documents placed in positions $1$ through $k$
of the answer pages returned by a search system, are categorical
data.
There is no total ordering of $k$-element SERPs, and unless we
get lucky and happen to have two SERPs that have a relationship
captured by a partial ordering, direct comparison of SERPs is not
possible.
The primary operation that is permitted on categorical data is
tabulation of occurrence frequencies, to determine the mode (most
frequently occurring value), or to determine the fraction of the data
observations that consist of some particular value, where in the case
of $k$-element binary SERPs a ``value'' is one of the $2^k$ distinct
bit-patterns that can arise.
Neither of those tabulations is helpful when seeking to determine
whether retrieval system $A$ is demonstrably better than retrieval
system $B$ when used for some specific type of search task.

Comparative evaluations of SERPs must thus be guided by pragmatic
requirements, and is usually carried out using a numeric
{\emph{effectiveness mapping}} from the categorical universe of
possible $k$-element SERPs to the real number line.\footnote{These
mappings have been previously referred to as effectiveness metrics,
evaluation metrics, and effectiveness measures.
To avoid ambiguity with the metric spaces that play an important role
in mathematics, and which imply properties not satisfied by sets of
SERP effectiveness scores, we make use here of the phrase
{\emph{effectiveness mapping}}.}
For example, the effectiveness mapping {\emph{reciprocal rank}}
({\metric{RR}}) assigns the score $1/d$ to all SERPs in which the
first $1$-bit is in rank position $d$, and assigns a score of zero to
the SERP $[0]^k$ containing $k$ zeros.

A wide range of other effectiveness mappings have also been proposed
over the years, each of which makes different assumptions about how
SERPs are valued relative to each other, based on different models of
user behavior.
These include precision (\Prec), average precision (\AP), normalized
discounted cumulative gain (\NDCG), rank-biased precision ($\RBP{}$),
plus others that the reader will be familiar with.
Each of these is a numeric mapping that (in ``@$k$'' evaluations)
maps the categorical set of $2^k$ different $k$-item binary vectors
to a value on the real line, within the range (by convention) from
zero to one inclusive.

The tension between SERPs being categorical data and the desire for
numeric comparisons between SERPs then leads to a critical question:
\begin{quote}
{\emph{Can any such categorical-to-numeric effectiveness mapping be
used, provided it is argued (by the person planning to employ it) to
be a plausible surrogate for SERP ``benefit to user'' in the context
of the given search scenario and the assumed user demographic?
And if not, what restrictions limit the choice of
effectiveness mappings?}}
\end{quote}

That question has received substantial attention.
In range of papers though the last few years one group of authors has
presented an opinion that such mappings may only be used if the range
of the mapping (that is, the set of numeric values that emerges when
the mapping is applied to every one of the $2^k$ SERPs of $k$-items)
generates a set of points that are equi-spaced on the real line
{\citep{fuhr17forum,fflp19iirw,ffp19ieeekade,ffl20irj,fff21ieeeaccess}}.
Examples detailing the implications of that belief appear in
Section~\ref{sec-cleo}.

{\citet{sakai20forum}} and the current author
{\citep{ieeeaccess22moffat}} have argued the contrary -- that any
categorical-to-numeric mapping may be used in SERP evaluation,
provided that it is motivated as reflecting the way in which users
are consuming those SERPs.
We believe that mappings should be chosen so as to plausibly capture
some external assumption in regard to the measurement scenario being
pursued, but apart from that connection do not require any intrinsic
properties of their own.
For example, one possible external assumption might be that users
place $d$ times more value on having the first relevant document at
rank $1$ than they do on having the first relevant document at rank
$d$; a relationship that then leads directly to {\RR}.

Each other effectiveness mapping similarly corresponds to a model of
user behavior, meaning that the score associated with each SERP
captures -- or at least, approximates -- the extent to which it is of
benefit to the user that issued the query.
From this point of view the user model can be thought of as coming
first, and then once the factors that govern the user model have been
settled, the desired categorical-to-numeric effectiveness mapping
will emerge {\citep{mmta22sigir}}.

The debate between these two points of view has now percolated into
the guidelines provided to conference referees, creating new threads
of contention.
For example, in November 2022 the SPC members for ECIR 2023 were
requested to vet their pools of papers against a set of guidelines
that included the (disputed) statement ``{\emph{RR is widely used but
difficult to interpret as interval scale, which would be a
requirement for computing the arithmetic averages}}'', so that the PC
Chairs could ``{\emph{instruct all authors of every accepted paper to
check and correct if it applies}}''.\footnote{The email in question
is dated 25 November 2022 and is from EasyChair, over the names of
the three PC Chairs.
The email refers to the guidance as coming from ``{\emph{our
`official' ECIR Methodology Chair}}'', but no such role is listed at
{\url{https://ecir2023.org/organisers.html}}, and no author is listed
in the document either, meaning that the precise origin of the
instruction
is unknown.
At time of writing these ``guidelines'' remain available at
{\url{https://docs.google.com/document/d/1taETBOHpnWCAOps-rVDV-j_ecUJS0U7S}}.
}

Most recently, in December 2022, the first group of ``{\RR} should be
avoided'' authors has again argued that measurement is sound only if
the numeric mapping involved employs a set of available target points
that are uniformly spaced on the real line {\citep{fff22arxiv}}.

In short, one group of authors in the IR community is strongly
opposed to {\RR} (and several other commonly used effectiveness
mappings; {\RR} is simply an exemplar that is used here in order to
allow a focused discussion) as an experimental tool, and has been
seeking to enforce that view across the community.
In opposition, a second group of authors believes that there is no
basis for such a restrictive view, and that researchers are free to
select effectiveness mappings that suit their experimental context,
provided only that they are willing to argue the {\emph{why}} of
their choice.

I'm firmly of that second opinion.
My goal in writing this essay is to reiterate my previous ``you may
use {\RR} if you wish'' position, and to provide further information
to colleagues who have been puzzled (or, in the case of negative
reviewing decisions, feel they have been adversely affected) by this
debate.
 \section{A Brief Quiz}
\label{sec-cleo}

The reader is invited to consider the following sequence of
questions, imagining perhaps that they are completing a ``how data
literate are you?''\ quiz in the sealed center section of some
research journal.
In all cases the word ``average'' refers to the arithmetic mean.

\question{1}{What is the average of the following eight
numbers? $$700, 700, 750, 750, 750, 850, 1000, 1000$$}{$812.5$} 

\question{2}{What is the average of the following eight monetary
amounts?
$$\$700, \$700, \$750, \$750, \$750, \$850, \$1000, \$1000$$}{\$$812.50$}

\question{3}{What is the average of the eight amounts listed in
Question~2 if it is additionally known that they are the weekly
salaries of eight employees?\\}{\$$812.50$} 

\question{4}{What is the average of the eight weekly salaries if it
is additionally known that the eight employees are two junior
professors, three assistant professors, one associate professor, and
two full professors?\\}{\$$812.50$} 

\question{5}{What is the average of the eight professor's salaries if
it is additionally known that all junior professors are paid $\$700$
per week, all assistant professors are paid $\$750$ per week, all
associate professors are paid $\$850$ per week, and all full
professors are paid $\$1000$ per week?\\}{$\$812.50$} 

\question{6}{What is the average of the eight professor's salaries if
it is additionally known that there are no other professorial job
titles, and no other professorial salary levels?\\}{$\$812.50$}
\medskip

\noindent
I imagine that the majority of readers gave a clear {\ansa} response
to Question~1, and then continued to respond {\ansa} even as further
information was provided about what it was that was being averaged.
Those six {\ansa}s are also the answers that I'm arguing for in this
essay.
Averaging numeric and monetary amounts is a routine operation, and we
would probably also feel equally comfortable (for example) if were we
to be asked to sum the same sets of values.

That comfort arises because none of Questions~1 to~6 are asking about
``average professors''.
The question sequence starts by asking about the average of a given
set of numbers without mentioning professors at all, and ends by
asking about the average salary of a group of professors, with the
set of {\emph{observations}} (the professors' ranks) being mapped to
a set of numeric quantities (the salaries corresponding to those
observations) in the form of numbers that have ``\$'' signs attached.
If we were to be asked about the ``average professor'', we would
shrug our shoulders, and quite properly respond with ``(\emph{b}) it
is not appropriate to compute an average''.
We would recognize that there is no basis for computing the average
of ``two junior professors, three assistant professors, one associate
professor, and two full professors'', because that is a dataset of
{\emph{ordinal}} observations.\footnote{However it {\emph{is}}
permissible to enquire about the median of ordinal observations.
For that same set of eight observations, the median professor is an
``assistant professor''; and the ``median professor's salary'' is the
mapped salary value associated with an assistant professor.}

In opposition to that view of when averaging may be applied, and as
part of a direct response to my previous paper
{\citep{ieeeaccess22moffat}} in which I introduced this example of
professorial salaries, {\citet[page~12]{fff22arxiv}} write:
\begin{quote}
	``{\emph{the examples by Moffat on professor salaries}}
	[\dots] {\emph{are not interval scales, as a consequence
	means cannot be computed}}''
\end{quote}
Followup emails with Nicola Ferro\footnote{My thanks to Nicola for
being willing to engage in this conversation.}
in November 2023 have confirmed that it is his belief that Question~6
should be answered {\ansb}, because (in his words) ``{\emph{I
understand the numbers do not come from an interval scale}}''.
Nicola also argues that Questions~1 to~5 must be answered {\ansc},
because their formulation doesn't specify (or doesn't specify in
sufficient detail) what the numbers represent.

To tease out the implications of Nicola's assertions, suppose that
$\UNI$ is the universe from which some given set of numeric
observations is being drawn.
For example, in the case of the professors' salaries,
$\Uprof=\{\$700,\$750,\$850,\$1000\}$.
Nicola's sequence of answers to Questions~1 to~6 can then be seen as
being equivalent to the belief that when $\nU=|\UNI|$ is finite and
if a dataset of observations drawn from $\UNI$ (with replacement) is
to be averaged, then $\UNI$ must have the form:
\begin{equation}
	{\UNI} = {\UFFF}(m, c, \nU) =
		\{ m\cdot k + c \mid k \in \var{range}(\nU) \} \, ,
	\label{eqn-universe}
\end{equation}
in which $m,c\in{\cal{R}}$; in which $\var{range}(\nU) =\{k\in {\cal{I}}\mid 0\le
k < \nU\}$; and where $\cal{R}$ and $\cal{I}$ are
the sets of real values and integer values respectively.
Note the use of equality in Equation~\ref{eqn-universe}.
Nicola's assertion is that not only must every value in $\UNI$ be
expressible as a real-valued linear transformation of a corresponding
integer, but also that {\emph{every such equi-interval mapped value
over a dense range of integers $0\le k<u$ must be available in $\UNI$}}.
That is, Nicola believes that despite the professors' salaries all
being multiples of $m=\$50$, a dataset of observations over $\Uprof$
may {\emph{not}} be averaged, because the equi-spaced salaries
$\$800$, $\$900$, and $\$950$ are not available; with
$\Uprof\not=\UFFF(m,c,\nU)$ for every combination of $m$, $c$, and
$\nU$.

In particular, in Question~5 there is still uncertainty as to what
values might be possible as salaries and the possibility that
$\Uprof$ might yet satisfy Equation~\ref{eqn-universe}, which is why
Nicola's answer to Question~5 is {\ansc}; but that uncertainty is
dispelled by the additional information provided in Question~6, thus
leading to the switch to answer {\ansb}.
Pursuing the same argument one step further, if the three extra
options $\$800$, $\$900$, and $\$950$ are also ``available'' as
weekly salaries (for example, because a small number of year-long
merit-based bonus payments of either $\$50$ or $\$100$ per week are
determined at the start of each year) but happen to not be visible in
this current set of observations (because none of the current
professors are being paid a bonus), then the averaging in Question~6
{\emph{would}} (presumably) be permitted, because the salary universe
would now be equal to $\UFFF(\$50,\$750,9)$.

My main point in introducing the professors' salaries was that
similar questions can be posed in connection with SERP effectiveness
scores {\citep{ieeeaccess22moffat}}.
Let's continue the quiz.

\question{7}{What is the average of the following eight
numbers?
$$0, \frac{1}{3}, \frac{1}{3}, \frac{1}{3}, \frac{1}{2}, \frac{1}{2},
1, 1$$}{$0.5$} 

\question{8}{What is the average of the eight numbers in Question~7
if it is additionally known that they are effectiveness scores
associated with eight SERPs of length $k=3$ via the use of a
categorical-to-numeric effectiveness mapping?\\}{$0.5$} 

\question{9}{What is the average of the eight mapped SERP scores in
Question~8 if it is additionally known that there is one SERP with no
relevant document by depth $k=3$, three that have their first
relevant document at depth three, two that have their first relevant
document at depth two, and two that have their first relevant
document at depth one?\\}{$0.5$} 

\question{10}{What is the average of the eight SERP scores in
Question~9 if it is additionally known that SERPs with no relevant
document by depth $k=3$ are always assigned a score of $0$ and that
SERPs with the first relevant document at depth $d$ are always
assigned a scores of $1/d$; and thus that there are no other SERP
scores available when $k=3$?\\}{$0.5$} 
\bigskip

\noindent
In response to this second set of questions, I again argue that
{\ansa} is the appropriate answer throughout.
My basis for that statement remains that it is the {\emph{numbers}}
that are being averaged, and that just as Questions~1 to~6 were not
attempting to compute an average professor, {\emph{there is no
suggestion at all in Questions~7 to~10 that SERPs are being
averaged}}.
Search engine result pages are {\emph{categorical}} data, even when
abstracted as $k$-vectors over $\{0, 1\}$, and hence may not be
averaged, just as the professors' ranks (ordinal data) may not be
averaged.
Furthermore, it isn't even possible to compute the median SERP across
a set of observations, since SERPs {\emph{do not possess a total
order}}.
As a simple example to support that claim, consider the two
$5$-element SERPs $S_1=[1,1,0,0,0]$ and $S_2=[0,0,1,1,1]$, in which
the document order is from left to right.
There is no innate ordering possible between these two SERPs (one has
more early relevance, the other has more total relevance) and
different effectiveness mappings are free to assign their scores in
either order.
In particular, {\RR} prefers the first of those two SERPs, with
${\RR}(S_1) = 1.0 > 0.3333 = {\RR}(S_2)$; whereas ${\Prec}@5$ prefers
the second SERP, with ${\Prec}@5(S_2)=0.6>0.4={\Prec}@5(S_1)$.

If any of us were to be provided with the set of SERPs that was
observed in an IR experiment and asked to calculate either the
average or the median of the set, we would have no option but to
shrug and quite properly answer ``(\emph{b}) it is not appropriate''.
It is the {\emph{mapped SERP scores}} that are being averaged in
Questions~8 to~10, not the SERPs themselves.
Nor is there any suggestion at all that the effectiveness mapping can
or may be inverted once that average score has been computed, to
determine a SERP that corresponds to it.
The effectiveness mapping is many-to-one, and inversion is neither
possible nor sensible.

In opposition to my view of Questions~7 to~10, Nicola answers {\ansc}
to Question~7, and then {\ansb} to Questions~8 through~10.
The explanation he provides for this response is that from Question~8
onwards there is explicit mention of a categorical-to-numeric
mapping, implying (in his opinion) that the average may not be
computed.
Nor does the universe of possible values $\Urr=\{0,1/3,1/2,1\}$ that
is employed in these questions fit the form required by
Equation~\ref{eqn-universe}, providing (presumably) another reason
for Nicola to answer {\ansb} to Question~10.

\begin{figure}[p]
\centering
\includegraphics[width=120mm,trim=60mm 65mm 60mm 40mm,clip]{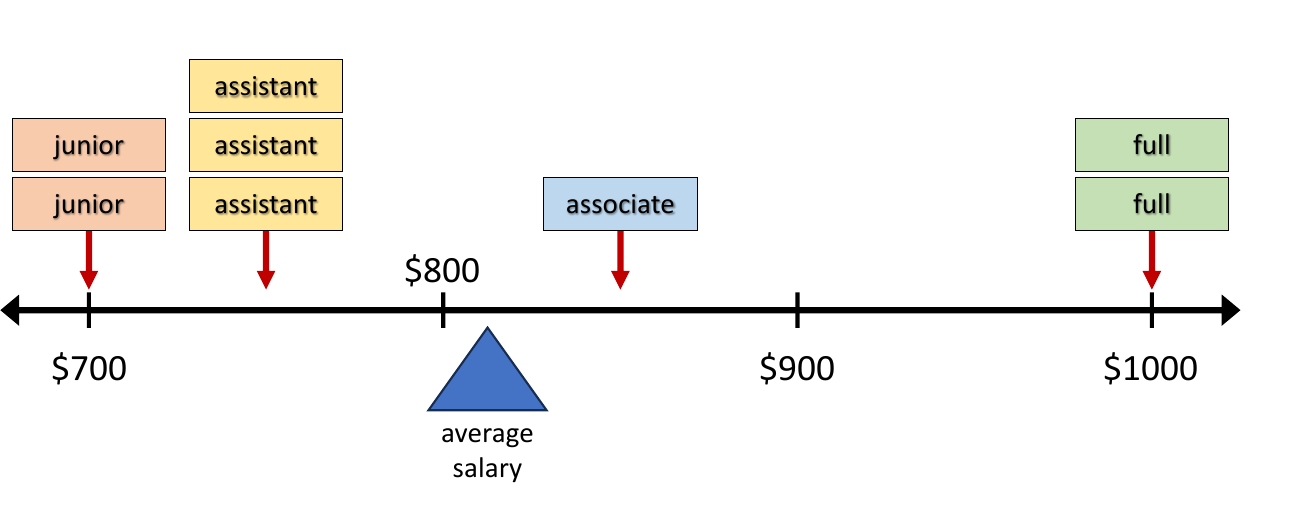}\\
{\small (a) Averaging eight professorial salaries.}\\[2ex]
\includegraphics[width=120mm,trim=60mm 65mm 60mm 40mm,clip]{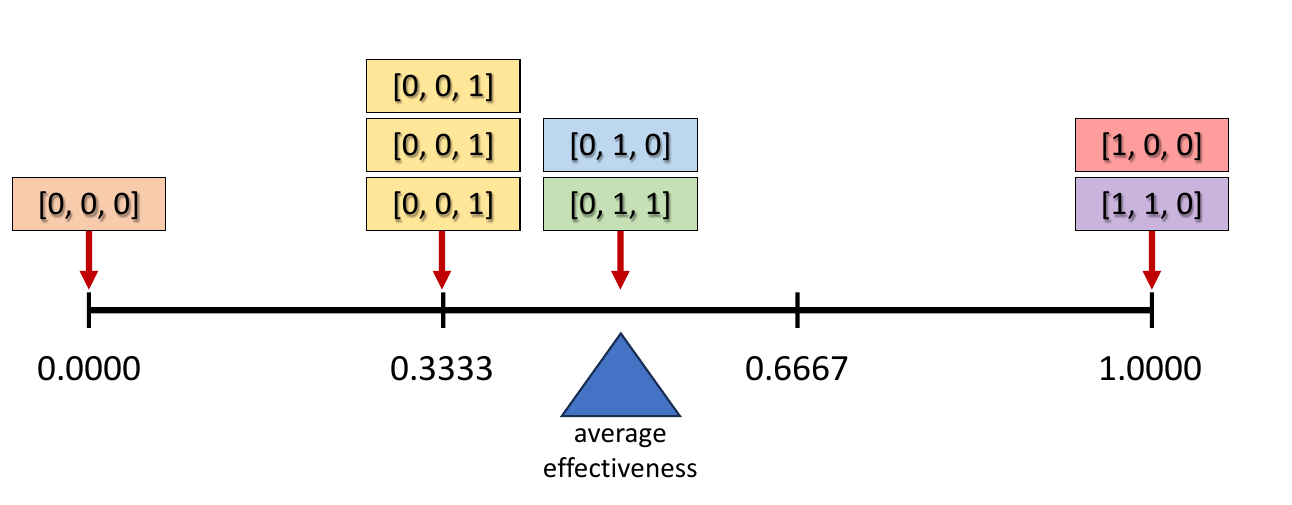}\\
{\small (b) Averaging eight ${\RR}@3$ effectiveness scores.}\\
\includegraphics[width=120mm,trim=60mm 65mm 60mm 15mm,clip]{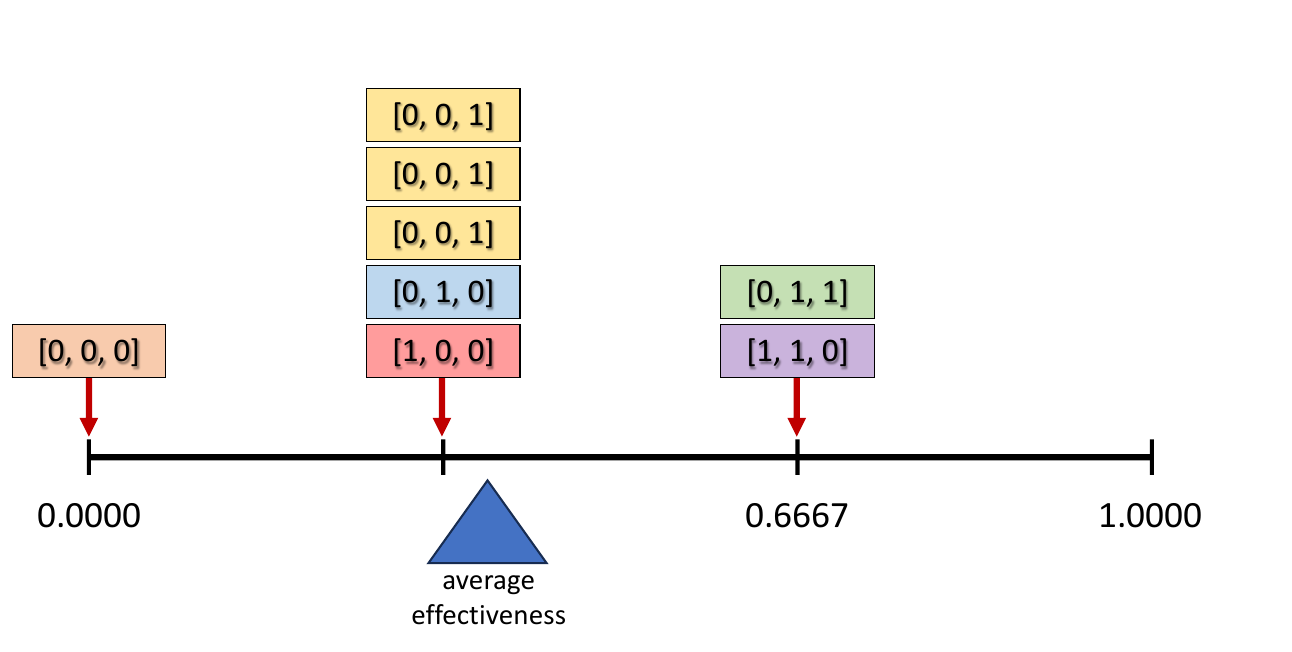}\\
{\small (c) Averaging eight ${\Prec}@3$ effectiveness scores.}\\
\caption{Averaging sets of values: (a) the four professorial salary
levels, and the distribution of the eight observations assumed in
Questions~1 to~6; (b) the four ${\RR}@3$ scores possible for SERPs of
length $k=3$, and the distribution of the eight observations assumed
in Questions~7 to~10; and (c) the same eight SERPs, but now assessed
using the effectiveness mapping ${\Prec}@3$.
\label{fig-serp-salaries}}
\end{figure}

Figure~\ref{fig-serp-salaries} illustrates the situations embedded in
the two sequences of questions.
Each part shows a {\emph{ruler}}, and a set of associated
{\emph{tick-points}} with their corresponding {\emph{numeric
labels}}.
As well, the mapped values corresponding to the sets of eight data
observations are shown.
In Figure~\ref{fig-serp-salaries}(a) the salaries of the eight
professors are depicted, along with their average.
The average salary of the eight professors has a clear and tangible
physical interpretation -- it is the fulcrum point at which the data
items become symmetric in a weighted sense, to bring the ``balance
beam'' into equilibrium.
That is, ``averaging'' has a compelling (and useful, if you like
see-saws!)\ analogy in the physical world, namely the identification
of fulcrums for balance beams.
In this physical correspondence there is no requirement whatsoever
that the observations being averaged must be located at the ruler's
equi-interval labeled tick-points, and the mapping being used can
place the items anywhere.
Nor is there any requirement that every marked tick-point on the
ruler must be capable of having an item placed on it.

Figure~\ref{fig-serp-salaries}(b) provides a similar picture in
connection with Questions~7 to~10, taking (for concreteness) the
eight SERPs to be
$[0,0,0]$,
$[0,0,1]$,
$[0,0,1]$,
$[0,0,1]$,
$[0,1,0]$,
$[0,1,1]$,
$[1,0,0]$, and
$[1,1,0]$.\footnote{A set of SERPs can always be placed into
lexicographic order.
Lexicographic ordering happens to agree with the ordering induced by
{\RR}, but, in general, mapped score orderings are not preserved.
As a simple example of that, note that
${\Prec}@3([0,1,1])\not\le{\Prec}@3([1,0,0])$.}
A balance point can always be identified, regardless of where on the
balance beam the data observations are placed, and can be identified
even if some tick-points are unavailable to host ``weights''.

In Figure~\ref{fig-serp-salaries}(a) the fact that the underlying
observations are ordinal, and that there are just four distinct
observations possible, means that if a different salary mapping is
applied to this dataset the stacks might shift sideways to new points
on the number line, but cannot change their height, nor ``swap over''
and shift past each other.
On the other hand, in Figure~\ref{fig-serp-salaries}(b) the
underlying observations are categorical, and there are $2^3=8$
different possibilities, six of which appear in this dataset.
A different $k=3$ effectiveness mapping might thus give rise to as
many as six different stacks for this set of observations, and as
many as eight stacks in general.
Moreover, the stacks for $[1,0,0]$ and $[0,1,1]$ might appear in
either order along the number line, whereas (making reasonable
assumptions about the partial ordering on SERPs) the other six stacks
may ``slide'' and perhaps even ``combine'', but may not ``swap
over''.
See {\citet{arxiv23mm}} for discussion of the extent to which this
``freedom to swap'' is available.

Figure~\ref{fig-serp-salaries}(c) gives one such alternative mapping
for the same set of SERPs, now using ${\Prec}@3$.
Because $|\Uprec|=4$ too, there are still at most four stacks, but
they have different memberships, a consequence of the greater freedom
associated with categorical data.
In this second configuration we have $\Uprec = \{0, 1/3, 2/3, 1\}
=\UFFF(1/3, 0, 4)$, a mapping that is compliant with
Equation~\ref{eqn-universe}.
However Nicola hedges when asked about this specific situation,
writing [the answer] ``{\emph{is (b) because you explicitly say it is
a categorical scale; however, if I do not use the information about
the categorical scale, I would say (a) because {\dots}
the objects (runs) are
equi-spaced}}''.
Needless to say, I believe that the ${\Prec}@3$ scores may be averaged
in exactly the same way that the ${\RR}@3$ scores may be.
In both cases the average continues to have a direct physical
interpretation, and the fact that ${\RR}@3$ and ${\Prec}@3$ stack the
SERPs differently on the number line, assign different scores, and
yield different average scores is of no consequence, because
{\emph{they employ different effectiveness mappings, and are thus
measuring different things}}.
And neither of them is claimed to be ``averaging SERPs''.

In terms of the extremes of the set of mapped values, $\min(\UNI)$
and $\max(\UNI)$, most researchers would agree that a SERP with no
relevance, $[0]^k$, is a natural {\emph{bottom}} element to the SERP
categories, and is thus normally mapped to zero.
Other SERPs might also be assigned a score of zero, but $[0]^k$
{\emph{will}} have a score of zero.
Similarly, a SERP that is all relevant, $[1]^k$ is a natural
{\emph{top}} element to the SERP categories, and is thus normally
mapped to the maximum score (typically $1.0$, or approaching $1.0$ in
the limit as $k$ increases).
Other SERPs might also be assigned a maximal score, but $[1]^k$
{\emph{will}} have a maximal score.
That is, despite SERPs not possessing a total order, there is a
partial order, with {\emph{top}} and {\emph{bottom}} elements that
bookend the range of values generated by any plausible
categorical-to-numeric mapping that might be employed in IR
evaluation.

To summarize the arguments presented in this section: SERPs are
categorical data, and hence the only directly permissible operation
is tabulation of frequencies, which is unhelpful in IR
experimentation.
We may not even ask about the median of a set of SERPs.
But each plausible model of the interactions that take place as users
consume SERPs gives rise to a corresponding categorical-to-numeric
effectiveness mapping.
All such mappings share a common natural zero point, and give rise to
up to $2^k$ different mapped scores in a many-to-one manner.
The set of SERP scores derived from each such mapping may then be
legitimately averaged without any implication that the SERPs
themselves are being averaged, nor that the mean score may be
converted back to a SERP.
And the mapping process is valid even if the numeric values that the
universe of $2^k$ different SERPs transform to are not evenly-spaced
on the number line.

Supporting that position is the clear fact that in the physical
analogy expressed via Figure~\ref{fig-serp-salaries} there is no
requirement that the objects placed on the balance beam be located
solely at the equi-spaced labeled tick-points on the ruler; and nor
is there any constraint that renders the averaging process improper
if there are labeled tick-points at which none of the weights will
ever be placed.
That is, there is a clear physical analogy that supports the
correctness of answer {\ansa} to all ten questions in the data
survey.

Of course, if the mapping changes and (in the context of
Figure~\ref{fig-serp-salaries}) the target points of some or all of
the red arrows shift, then the fulcrum point is likely to also need
to shift.
But that is perfectly sensible -- the average is determined
{\emph{after}} the mapping is applied to the set of observations, and
there is {\emph{no suggestion at all}} being made here that the
average is (or even should be) invariant to the
categorical-to-numeric effectiveness mapping that has been chosen.
Such invariance is simply not possible in connection with categorical
(SERPs) or even ordinal (professorial ranks) data.
 \section{Interval Scales Versus Equi-Interval Scales}
\label{sec-interval}

Presuming that the reader has absorbed the arguments that were
presented in Section~\ref{sec-cleo}, they may now be somewhat
puzzled.
If, as I have asserted, SERPs are categorical data, but may be mapped
to the real number line via some defended choice of non-reversible
categorical-to-numeric effectiveness mapping, and if all plausible
mappings on SERPs have a defined zero point, then what are interval
scales, and why do they need to be considered?

Let's start with what an interval scale is.
The best example of interval scale data is the one I noted in my
previous paper {\citep{ieeeaccess22moffat}}, Unix timestamps.
The Unix operating system measures time as elapsed seconds, starting
at 0:00:00am on 1 January 1970.
The current time a few minutes ago was $1694942015$, and sixty
seconds prior it was $1694941955$.
In an interval scale measurement system the actual raw measurements
may not be especially informative (for example, the ten-digit numbers
provided in the previous sentence) unless the zero point is also of
significance to the observer.
But {\emph{differences}} between observations provide information;
furthermore, {\emph{ratios of differences}} may be computed and also
provide information.
For example (dropping six leading digits, to make the numbers
manageable), $2015 - 1955 = 60$ and represents an interval (in
seconds) that can be correctly stated as being $1.5$ times longer
than the interval between $2310$ and~$2350$.

Unix times are normally thought of as being integers, but from a
measurement point of view there is no requirement that they be
restricted to integral values, and fractional values are also
completely fine.
Thus the interval from $2380.5$ to $2396.4$ is $1.5$ times longer
than the interval from $2320.5$ to $2331.1$, because
$(2396.4-2380.5)/(2331.1-2320.5) = 1.5$.
Moreover, if an alternative reference time was chosen, for example,
2:22:22am on December 22, 2022, all of the raw measurements would
change, but the differences, and ratios, between them, would be
invariant.
And if the measurement units altered from seconds to (say) tenths of
seconds, both the raw measurements and their differences would
change, but the ratios between the differences would again remain
constant.
This example highlights the essential property of interval scale
observations, namely that the zero point may be arbitrary, but that
ratios of differences between pairs of observations hold constant,
regardless of where the zero is placed and regardless of what the
measurement unit is.
When measuring time at the granularity of minutes or seconds or hours
or days (but not years, because of the irregularity introduced by
leap years) two different measurement systems will have this property
of being consistent in terms of ratios of differences.
We handle these scale and origin transformations dozens of times
every day, and (as one further example) know that a meeting scheduled
for 3:00pm--4:30pm tomorrow is $1.5$ times longer than one that is
scheduled for 10:15am--11:15am today.

In an interval scale each unit change in the measurement must have
the same interpretation, regardless of where in the scale it occurs.
Those unit changes can be denoted by equally-spaced tick marks on a
ruler and then be labeled with corresponding equally-spaced numeric
measurements, as was shown in the examples provided in
Figure~\ref{fig-serp-salaries}.
But must a given set of observations (professorial ranks mapped to
salaries, or SERPs mapped to effectiveness scores, or seconds since
0:00:00am on January 1, 1970) be required to {\emph{only}} fall on
the equi-interval tick marks of some specific ruler in order for the
measurement (and more particularly, an average of measurements) to be
possible or valid?
No!
The observations can fall anywhere, and we can still measure them,
and then still average those measurements.
Indeed, the constancy of ratios of differences that is guaranteed by
an interval scale means that we may infer additional ticks on the
ruler {\emph{anywhere that they might be helpful}}.
If the salary ruler is labeled and ticked every $\$100$, we may infer
tick points at $\$50$ intervals as well, by taking ratios at
intervals that are $0.5$ long relative to the original tick
intervals.\footnote{And that is exactly the process that was used to
place some of the elements of Figure~\ref{fig-serp-salaries}.}
Or we can infer tick marks every $\$10$, by taking ratios at
intervals that are $0.1$ long relative to the original tick
intervals.
Indeed, we can infer a tick that corresponds to the exact position on
the ruler of {\emph{any object being measured}}, including a salary
of $\$765.43$ per week, or an {\RR} score of $1/7\approx0.1429$, or a
time that is $314.159$ seconds after 0:00:00am on January 1 1970.

We are not changing the scale when we add such ``pro-rata'' ticks to
the ruler, nor are we changing the previous equally-spaced unit
labels.
We are simply employing the existing ruler, and our knowledge that it
measures an interval scale (and thus that ratios of differences have
a valid interpretation) to strategically add further ticks, so as to
measure objects whose size is not one of the original tick points on
the ruler.
I'll say it again: {\emph{new tick-points can be placed anywhere on
the ruler, and once they have been labeled in a proportionate manner,
are first-class citizens, possessing the same rights, privileges, and
responsibilities as the original equi-spaced ticks.}}
An interval scale does {\emph{not}} require that all valid
measurements be at the places on the ruler where the original
evenly-spaced unit tick-marks arose.
Nor does it require that every original tick-point be attainable as
an observation.

In contrast to this flexible interpretation of interval scale
measurement, {\citet{fuhr17forum}} and {\citet{fff21ieeeaccess}}
espouse the more rigid requirement that is documented in
Equation~\ref{eqn-universe}.
They require that the tick-marks and their labels must only be at
uniform intervals on the ruler; that every mapped observation must
exactly align with one of those equi-spaced tick-marks; and that
every tick-mark that appears on the ruler must be capable of being
such a measurement point.

In particular, {\citet{fff22arxiv}} assert that the professorial
salaries shown in Figure~\ref{fig-serp-salaries}(a) are not
``interval scale'' because the red arrows denoting the range of the
mapping are not equi-spaced on the number line -- see the quotes
already provided in Section~\ref{sec-cleo}.
{\fff}\ claim support for their interpretation from the original work
in this area, that of {\citet{stevens46}}, who in connection with
interval scales wrote ``{\emph{equal intervals of temperature are
scaled off by noting equal volumes of expansion}}'', and ``{\emph{to
devise operations for equalizing the units of the scales}}''.
But in my opinion both of those quotes refer to the starting
configuration of the ruler, and the overarching question that
determines the scale of measurement: if the original tick-points are
equi-spaced on the ruler, then can equi-spaced numeric labels be
inferred, and if so, are those labels equi-spaced numerically?
I do {\emph{not}} believe that Stevens' comments imply a further
mandate that every data observation must fall at an equi-spaced
tick-point on the ruler; and nor that they imply that every
equi-spaced label on the original ruler must be available as an
observation value.

In my opinion the three rulers shown in
Figure~\ref{fig-serp-salaries} {\emph{do}} all have the property
underlying Stevens' words: tick-points that are the same ``distance''
apart have labels that are the same ``value'' apart.
Some of the measurement values do not align with the currently-drawn
tick-points, that is correct.
But as has already been discussed, the first two rulers can be
proportionately ``sub-tick-pointed'' to {\emph{any required level of
detail}} (twenty decimal places if necessary, exceeding the
resolution of even {\sf{double}} arithmetic) so as to carry out the
required measurements.
Provided that ratios of differences are preserved, {\emph{we can put
ticks and labels, and hence allow measurement, anywhere we want on
the ruler}}.
And, similarly, {\emph{we can remove from the ruler any tick points
and their corresponding labels that will not be needed when
determining the mapped measurements derived from a set of
observations}}.

To make this distinction concrete, I return to the professor's
salaries and the restrictions implied by Equation~\ref{eqn-universe}.
{\fff}\ require that if an ordinal-to-numeric salary mapping is to
permit averaging, then either equi-interval salaries of $\$700$,
$\$800$, $\$900$, and $\$1000$ (or similar) must be used; or several
new grades of professor need to be introduced, to allow salaries at
all other multiples of $\$50$ between $\$700$ and $\$1000$; or some
combination between these two.
That is, {\fff}\ regard mapped salary levels of $\$700$, $\$800$,
$\$900$, and $\$1000$ as being ``averagable'' given a set of
``professorial job title'' observations; and (would presumably) also
regard mapped salary levels of $\$700$, $\$801$, $\$902$, and
$\$1003$ as being ``averagable''.
But they regard averaging as being inappropriate for the salary
mapping that led to Question~6 (namely, $\$700$, $\$750$, $\$850$,
and $\$1000$), even though the ruler could equally have been ticked
and labeled at intervals of $\$50$; and for the same reason would, I
presume, regard a salary mapping to the values $\$701$, $\$800$,
$\$900$, and $\$1000$ to also be non-averagable, even though the
ruler could equally have been ticked and labeled at intervals of~$\$1$.

In short, I'm bemused by the position that has been advocated by
{\fff}, and I firmly believe that the extraneous ``interval scale''
constraints (captured via Equation~\ref{eqn-universe}, in line with
Nicola Ferro's answers to the quiz questions) are neither required
nor helpful.
Certainly, in the balance beam physical analogy
(Figure~\ref{fig-serp-salaries}) there is no requirement that the
weights on the beam be placed solely at a set of equi-spaced labeled
tick-points on the ruler in order for the fulcrum point to be
determinable; and nor is the averaging process rendered improper by
the presence of labeled tick-points on the ruler at which none of the
weights can ever be placed.

All of which means that (in my opinion) effectiveness mappings like
{\RR}, {\AP}, {\NDCG}, $\RBP{0.8}$ and so on may be used with a clear
conscience -- provided, of course, that the user model behind the
selected mapping can be argued as matching the task type and the
behavior of the people carrying out that task.
Of course, the things being averaged must be numbers.
As I noted last year {\citep{ieeeaccess22moffat}}, we can't average
international dialing prefixes such as ``+1'', ``+61'', and ``+40''.
And nor can we average food additive codes such as ``414'', ``950'',
and ``260''.
In both of those examples the observations are categorical, just like
SERPs.
We may average the numbers that result from the application of a
suitable (and argued for) categorical-to-numeric mapping if we have
reason to do so, but the observations themselves cannot be averaged.
 \section{Ratio Scales}
\label{sec-ratio}

My next point suggests that there really is no need to be concerned
about what is and what is not an interval scale, and it is this:
{\emph{effectiveness mappings usually have a defined zero point}}, and
hence {\emph{effectiveness scores are ratio scale data}}.
In interval scale data the location of the zero point is arbitrary,
meaning that the computation of ratios of measurement differences is
permitted (because differences are origin independent), but that the
computation of ratios of actual measurements is not appropriate.
But in ratio scale data the fixing of the zero point means that
ratios may be computed on score {\emph{values}} as well as on score
{\emph{differences}}.
For example, it is not appropriate to state that $20^{\circ}$ is
``twice as hot as'' $10^{\circ}$ if we are measuring temperature
using Celsius or Fahrenheit, both of which interpret ``temperature''
as being interval scale observations.
In those two measurement systems we are only permitted to say that
$20^{\circ}$ is ``$10^{\circ}$ hotter'' than $10^{\circ}$, taking
differences to do so.
But we are permitted to say that $20^{\circ}$ is ``twice as hot as''
$10^{\circ}$ if the temperatures are being measured in degrees
Kelvin, which has a fixed zero point, and is thus a ratio scale.
(Noting, of course, that $20^{\circ}$ Kelvin is a different
temperature to $20^{\circ}$ Fahrenheit, and that $20^{\circ}$ Celsius
is different again).

Salaries also have a defined and unambiguous zero point.
That means that in the context of the professorial salaries supposed
by Questions~1 to~6 (that is, in the context established by one given
ordinal-to-numeric mapping) it is both defensible and informative to
compute ratios and make statements such as ``at that university an
assistant professor earns three-quarters the salary of a full
professor''.
It is equally valid to compute a difference and write ``at that
university an associate professor earns $\$100$ more per week than
does an assistant professor''.
Neither of those statements is invariant to the mapping, and if the
salary scales change, or a different university is considered, the
relativities may well change too.
But {\emph{in the context of the specified mapping}}, such statements
are legitimate.

With those examples in hand, let's now turn back to
categorical-to-numeric effectiveness mappings.
It was noted in Section~\ref{sec-cleo} that the binary $k$-vector
$[0]^k$ is almost always mapped to a score of zero.
And zero does indeed have a fixed and identified meaning when
evaluating SERP effectiveness -- it denotes the absence of
usefulness.
Hence, if the effectiveness mapping being applied to a set of SERPs
is one that maps $[0]^k$ to zero, then the set of mapped
effectiveness scores constitutes {\emph{ratio scale}} data in the
context provided by that mapping.
And that means, for example, that we are permitted to state (for that
specified mapping) that an effectiveness score of $0.6$ is ``$1.5$
times larger'' than an effectiveness score of $0.4$, as well as being
permitted to state that it is ``$0.2$ bigger''.
But it is also important to note that such statements {\emph{do not
and must not}} be used to make claims such as ``if $M_1(S_1)$ is
$1.5$ times bigger than $M_1(S_2)$, then $M_2(S_1)$ will also be
$1.5$ times bigger than $M_2(S_2)$'', where
$S_1$ and $S_2$ are two SERPs, and $M_1(\cdot)$ and $M_2(\cdot)$ are
two effectiveness mappings.
As with the professor's salaries, conclusions are not, and never can
be, invariant to the mapping.

To summarize this argument: if the effectiveness mapping in use
assigns a score of zero to useless SERPs (that is, those that are
$[0]^k$), then discussion of interval measures is irrelevant, because
-- strictly within the context established by that effectiveness
mapping -- the (broader) power of ratio scales may be applied to the
set of SERP scores.
Nor are the categorical-to-numeric mappings ${\Prec}c@k$ and $\RBP{0.5}$
any different in this regard.
Yes, they do happen to have the property that (when there are $k$ or
more relevant documents in the collection) the mapped scores are a
dense equi-spaced set of points on the number line, satisfying
Equation~\ref{eqn-universe}.
But it is {\emph{still}} not possible to make claims such as ``if
${\Prec}c@k(S_1)$ is $1.5$ times bigger than ${\Prec}c@k(S_2)$, then
$\RBP{0.5}@k(S_1)$ will also be $1.5$ times bigger than
$\RBP{0.5}@k(S_2)$''.
As was noted in Section~\ref{sec-cleo}, SERPs themselves are
{\emph{categorical}} data, and it is simply impossible for there to
be any order-based, interval-based, or ratio-based invariance.
The bottom line here is that {\emph{each different effectiveness
mapping is measuring a different thing}}.

Finally, there is one more issue to address.
{\cite{fff21ieeeaccess}} proceed to describe the
{\emph{intervalization}} process that is summarized in
Table~\ref{tbl-intervalization}.
A conventional IR experiment can be thought of as following steps~0
to~3 of that experimental sketch, leading to sets of scores that, in
line with my arguments here, may then be averaged or compared in
other ways.
But having claimed that in general such sets of real-valued
effectiveness scores may {\emph{not}} be averaged, {\fff}\ propose
the ``amelioration'' shown at steps~4 and~5.
Their suggestion is that the scores be converted to ordinal integers,
by assigning a rank between $1$ (for the lowest possible score) and
$2^k$ (or less, depending on the effectiveness mapping, for the
highest possible score) to each possible numeric value that can arise
from the chosen categorical-to-numeric mapping when applied to the
universe of $2^k$ binary $k$-element SERP abstractions.
And that those score ranks, which as is noted in the table, are now
{\emph{ordinal}} data, be converted back to effectiveness scores by
dividing each rank by the number of distinct scores arising, that is,
by dividing by $u=|\UNI|$.

\begin{table}[t]
\centering
\newcommand{\tabenta}[1]{\begin{minipage}[t]{85mm}\small\raggedright #1\\~\end{minipage}}
\newcommand{\tabentb}[1]{\begin{minipage}[t]{65mm}\small\raggedright #1\\~\end{minipage}}

\begin{tabular}{cll}
\toprule
Step
	& Activity
		& Observations and measurement scale
\\
\midrule

0
&\tabenta{Design experiment; identify documents, topics, and queries;
select and justify retrieval effectiveness mapping.}
&\tabentb{---}
\\

1
&\tabenta{Carry out experiment, collect SERPs created by system(s)
in response to queries.}
&\tabentb{Document $k$-vectors, one per system per query;
{\emph{categorical}} data.}
\\

2
&\tabenta{Collect relevance judgments, then apply them to
the document $k$-vectors.}
&\tabentb{Binary $k$-vectors, one per system per query;
{\emph{categorical}} data.}
\\

3
&\tabenta{Apply effectiveness mapping chosen at step~0
(categorical-to-numeric)
to the binary $k$-vectors, with $[0]^k \rightarrow 0$.}
&\tabentb{Numeric scores, one per system per query;
{\emph{ratio}} data.}
\\

4
&\tabenta{Transform each mapped score to its rank position in the
sorted list of all possible scores possible for that mapping
when applied to binary $k$-vectors.}
&\tabentb{Score ranks relative to $\UNI$, the universe of possible
scores for the mapping chosen at step~0;
{\emph{ordinal}} data.}
\\

5
&\tabenta{Transform SERP ranks to the range $[0, 1]$ using equi-interval
ordinal-to-numeric mapping.}
&\tabentb{Mapped score ranks in the range $[0, 1]$;
\\{\emph{ratio}} data again.}
\\
\bottomrule
\end{tabular}
 \caption{Experimental structure suggested by
{\citet{fff21ieeeaccess}}.
Steps~0 to~3 are the standard IR experimental methodology; steps~4
and~5 are then added by
{\fff}
The commentary at the right is my analysis of the data that is
created by each step.
\label{tbl-intervalization}}
\end{table}

In this proposal if $M(\cdot)$ is the effectiveness mapping selected
in step~0 and applied in step~3, first the universe of all possible
mapped values $\UNI=\{M(S)\mid S\in[\{0,1\}]^{k}\}$ is computed; then
the cardinality of that set $u=|\UNI|$ is determined; and then each
SERP $S$ observed in the experiment is represented by a ``new'' score
$M'(\cdot)$ computed as
\[
	M'(S) = \frac{\var{rank}(M(S), \, \UNI)-1}{u-1} \, ,
\]
where $\var{rank}(s, \UNI)$ returns the integer value $|\{x \mid
x\in\UNI\wedge x\leq s\}|$; and where the ostensible goal is to
convert non-equi-interval observations to equi-interval observations
via a non-linear transformation from the real number line to the real
number line.\footnote{Which is not a permissible operation for either
interval scale data or ratio scale data.}
For example, in the case of the ${\RR}@3$ mapping that was depicted
in Figure~\ref{fig-serp-salaries}(b), we have $\Urr=\{0, 1/3, 1/2,
1\}$, and thus $u=4$; which in turn means that the adjusted mapping
$M'(\cdot)$ assigns each SERP $S$ one of $\UFFF(1/3, 0, 4)=\{0, 1/3,
2/3, 1\}$, ``sliding'' the stack of observations at $1/2$ in
Figure~\ref{fig-serp-salaries}(b) rightward, to sit at $2/3$.
This new effectiveness mapping could be used in an IR experiment, yes,
{\emph{provided it is argued for based on a plausible model of user
behavior}}.
But it no longer reflects the user model associated with ${\RR}@3$; it
measures SERP usefulness via a different user model.

The annotations in the right column in
Table~\ref{tbl-intervalization} make clear what, in my opinion, is
happening: step~3 creates ratio scale values that, by virtue of and
in the context of the user model associated the chosen effectiveness
mapping, are believed to represent user satisfaction with observed
SERPs.
Those ratio scale values are then being deliberately stripped of
their ratio-based relativities (step~4).
Instead, the scores are distorted into a new arrangement (step~5)
that possesses the same ordering, yes, but associates different
scores; thus undermining the interpretability of the experimental
results.

{\cite{fff21ieeeaccess}} further note that in certain cases
$M'(S)=c\cdot M(S)$ for a multiplicative constant $c>0$ and all
binary $k$-vectors $S$.
For example, provided there are at least $k$ relevant documents, both
${\Prec}c@k$ and $\RBP{0.5}@k$ (the $\RBP{}$ mapping, with parameter
$\phi=0.5$ and calculation limited to a SERP length of $k$) have this
property.
On the other hand, {\RR} does not, as is evident in the previous
example.
But even when this property holds for two mappings $M_1(\cdot)$ and
$M_2(\cdot)$, they will {\emph{still}} not be ratio-of-difference
invariant or ratio-invariant with respect to each other.
{\emph{Every categorical-to-numeric effectiveness mapping, whether
equi-interval or not, is measuring something different.}}

As I argued previously {\citep{ieeeaccess22moffat}}, I doubt the need
for the intervalization process.
Yes, it is another way of developing new effectiveness mappings that
might be argued for and then used; but no, those new mappings do not
possess any innate properties that make them, absent of any
discussion of users and their behavior, preferable to the mappings
they are derived from.
Intervalization most certainly does {\emph{not}} result in
order-based, ratio-of-difference-based, or ratio-based score
invariance across different effectiveness mappings.
That simply isn't possible, because the SERPs themselves are
{\emph{categorical data}}.

 \section{The Case For Reciprocal Rank}
\label{sec4-summary}

I'll admit to not being a great fan of {\RR} as a way of assessing
SERP usefulness.
It is a relatively blunt instrument, and for strongly top-weighted
evaluations I'd probably use $\RBP{0.5}$ as a measurement tool, so
that score ties are broken by taking into account the position of any
subsequent relevant documents after the first.
{\citet{diaz23arxiv}} has also recently considered how to add
sensitivity to {\RR} while retaining a strong focus on the head of
the SERP.

On the other hand, I'm also willing to accept that for some
applications {\RR} might be exactly the right way to assess user
perceptions of SERP usefulness.
For example, mention was made in Section~\ref{sec-intro} of search
tasks in which ``users place $d$ times more value on having the first
relevant document at rank $1$ than they do on having the first
relevant document at rank $d$'' -- which, by the way, is a statement
about relativities that directly corresponds with the use of ratio
scales.

Most importantly of all, I'd like my colleagues in the IR community
to be able to use {\RR} (or {\AP}, or {\NDCG}, or $\RBP{0.8}$, and so
on) in their evaluations if they wish to.
To satisfy me they need to provide a justification as to {\emph{why}}
they have chosen some particular effectiveness mapping, preferably
grounded in a discussion about user behavior; and in my opinion that
rationale is an important component of the experimental design
(step~0 in Table~\ref{tbl-intervalization}).
But once they have made that argument (and it has been accepted as
plausible) they should be able to report their results and present
their papers at conferences without being concerned that they may
have erred in some nebulous way that they don't understand, and
without attracting ``methodology'' criticisms that they don't
deserve.

Also worth noting is that recent work with Joel Mackenzie
{\citep{arxiv23mm}} has led to a proposal for comparing sets of SERPs
(in other words, comparing systems over a set of topics) via any
{\emph{innate pairwise SERP orderings}} that can be identified.
As was discussed in Section~\ref{sec-cleo}, there is no total
ordering of binary $k$-vectors, with the example $S_1=[1,1,0,0,0]$
versus $S_2=[0,0,1,1,1]$ used to defend the claim that SERPs are
categorical data.
But there are also many sub-chains and partial-orderings of SERPs.
For example, the two $5$-vectors $S_3=[1,0,1,1,0]$ and
$S_4=[0,1,0,0,1]$ probably {\emph{can}} be ordered -- at least for
all plausible effectiveness mappings $M(\cdot)$ -- as being
$M(S_3)\not< M(S_4)$, with $S_3$ possessing both more total
relevance, and also unambiguously earlier relevance.
That is, we cannot imagine any user-based model for SERP consumption
under which a user would find $S_4$ more useful than $S_3$.
This proposal is based on the ``Rule~1'' and ``Rule~2'' relationships
I have previously written about {\citep{ieeeaccess22moffat}}, and
leads to a technique we dub {\metric{IPSO}} for carrying out
mapping-agnostic system comparisons in some circumstances.
{\citet{df22sigir}} have also considered ``metric free'' comparisons.

Finally, at a more abstract level of experimental commentary, note
that {\citet{forum22zobel}} has recently pondered the degree to which
batch evaluations of the type considered in detail here should be
relied on as a way of assessing search performance; and discusses the
extent to which any numeric measurement can be employed as a reliable
surrogate for what is, fundamentally, a qualitative human-oriented
goal.

\subsubsection*{Acknowledgement}

This work was supported under the Australian Research Council’s
Discovery Projects funding scheme (project DP190101113).
A number of colleagues have contributed their thoughts in response to
drafts, including Joel Mackenzie, Justin Zobel, Paul Thomas, and
Tetsuya Sakai.

\begin{small}
\renewcommand{\bibsep}{2.5mm}

\end{small}

\end{document}